\documentclass[12pt]{article}

\usepackage{graphicx}  
\usepackage{dcolumn}   
\usepackage{bm}        
\usepackage{amssymb}   

\usepackage{color}

\def\edited#1{{\textcolor{blue}{#1}}}

\def\({\left(}
\def\){\right)}
\def\DS{\displaystyle}

\def\av#1{\Bigl\langle{#1}\Bigr\rangle}
\def\be#1{\begin{equation}\label{#1}}
\def\ee{\end{equation}}
\def\BM{\begin{bmatrix}}
\def\EM{\end{bmatrix}}

\def\[{\left[}
\def\]{\right]}
\def\av#1{\Bigl\langle{#1}\Bigr\rangle}

\newcommand{\Matr}[1]{\boldsymbol{{#1}}{}}
\def\edited#1{{#1}}
\def\eq#1{(\ref{#1})}

\def\paptit#1{#1}
\def\Xi{\Matr{\xi}{}}

\def\Gamma{\Matr{\gamma}{}}

\def\be#1{\begin{equation}\label{#1}}
\def\ee{\end{equation}}
\def\eq#1{(\ref{#1})}

\let\DS=\displaystyle

\def\({\left(}
\def\){\right)}

\let\DS     = \displaystyle

\def\<{\langle}
\def\>{\rangle}

\def\Xi{\Matr{\xi}{}}

\begin{document}



\title{Ballistic resonance and thermalization in Fermi-Pasta-Ulam-Tsingou chain \\at finite temperature}

\author{Vitaly A. Kuzkin and Anton M. Krivtsov \footnote{Peter the Great Saint Petersburg Polytechnical
		University; Institute for Problems in Mechanical Engineering RAS;
		e-mail: kuzkinva@gmail.com}}
\maketitle


\begin{abstract}
We study conversion of thermal energy to mechanical energy and vice versa
in $\alpha$-Fermi-Pasta-Ulam-Tsingou~(FPUT) chain with spatially sinusoidal profile of initial temperature.
We show analytically that coupling between macroscopic dynamics and
quasiballistic heat transport gives rise to mechanical vibrations with growing amplitude.
This new phenomenon is referred to as ``ballistic resonance''.
At large times, these mechanical vibrations decay monotonically,
and therefore the well-known FPUT recurrence paradox
occurring at zero temperature is eliminated at finite temperatures.
\end{abstract}
\maketitle


\section{Introduction}
Conversion of mechanical energy in solids to the thermal energy results in damping of mechanical vibrations,
whereas
thermal expansion and heat transport lead to the opposite conversion of thermal energy to the mechanical one.
In macroscopic continuum theories, the conversions are modeled by coupling between the equations of momentum
balance and of energy balance~(linear thermoviscoelasticity being
an example~\cite{Christensen Thermoviscoelasticity}).
However at micro- and nanoscale conventional macroscopic constitutive relations may be inapplicable.
For example, recent theoretical~\cite{Lepri 2003, Dhar 2008, Bonetto Lebowitz} and
experimental~\cite{Chang nanotubes, Johnson Temp Grat, Rogers Therm Grating,  Huberman graphite experiment,
Xu grap exp 2014}
studies  show that the Fourier law of heat conduction may be violated. In particular,
 ballistic regime of heat transport is observed~\cite{Hsiao ballistic, Anufriev nanowires}.

A well-known model of thermomechanical processes at micro- and nanoscale is the Fermi-Pasta-Ulam-Tsingou~(FPUT)
chain~\cite{FPU}. Despite the apparent simplicity of the model, analytical description of macroscopic
 thermoelasticity, heat transport, and energy conversion in the FPUT chain remains a serious challenge
 for theoreticians.

Several anomalies of thermomechanical behavior have been observed for the FPUT chain.
The heat transport in it is anomalous, i.e. the Fourier law
is violated~\cite{Das 2014, Das Spohn, Xiong}. The Maxwell-Cattaneo-Vernotte law also
fails to describe the heat transport in FPUT chains~\cite{Gendelman 2010 nonstat, Zhao Wang sin}.
Harmonic approximation allows one to derive equations~\cite{Krivtsov DAN 2015, Krivtsov 2019 ballistic}
and closed-form solutions~\cite{Kuzkin 2017 JPhys, Kuzkin 2019 CMAT ballistic} describing heat transport in chains.
However the question arises whether the temperature field, obtained in harmonic approximation, can be used
for estimation of thermoelastic effects~(e.g. excitation of mechanical vibrations due to thermal expansion).
We address this issue below.

Conversion of mechanical energy to thermal energy is even more challenging issue.
Studies of the conversion
have a long history, starting from the pioneering work of Fermi, Pasta, Ulam, and Tsingou~\cite{FPU},
where  initial conditions, corresponding to excitation of the first normal mode of the chain, were considered.
It was shown numerically that energy of this mode demonstrates almost periodic behavior,
 i.e. the system does not reach thermal equilibrium.
 In literature this phenomenon is referred to as the Fermi-Pasta-Ulam-Tsingou
 recurrence paradox~(see e.g.~\cite{Gallavotti 2008 FPU}).
 Recent advances in understanding of the paradox are summarized,
 e.g. in works~\cite{Gallavotti 2008 FPU, Berman 2005 FPU, Onorato 2015 FPU}.
 We note that in the original statement of the FPUT problem the chain has zero initial temperature.
  To the best of our knowledge, the conversion of mechanical energy to thermal energy at finite temperature
  has not been studied systematically.

Thus, in spite of significant progress in understanding of some particular thermomechanical phenomena
in the FPUT
chain~\cite{FPU, Das 2014, Das Spohn, Xiong, Gendelman 2010 nonstat, Zhao Wang sin, Gallavotti 2008 FPU,
 Berman 2005 FPU, Onorato 2015 FPU, Dmitriev PRB 2009},
a comprehensive  theory of macroscopic coupled thermoviscoelasticity for this system is yet to be developed.

In this paper, we report new thermomechanical phenomena observed in the $\alpha$-FPUT chain
with spatially sinusoidal profile of initial temperature.
Firstly, we show analytically that temperature oscillations,
caused by quasiballistic heat transport, and thermal expansion give rise to mechanical vibrations
with growing amplitude. This new phenomenon is referred to as ``ballistic resonance''.
Secondly, we show numerically that mechanical vibrations, excited by the ballistic resonance,
decay monotonically in time. Therefore at finite temperatures the FPUT recurrence paradox
is eliminated.

\section{Equations of motion and initial conditions}
%

We consider the $\alpha$-FPUT chain~\cite{FPU} consisting of~$N$ identical particles of mass~$m$, connected by nonlinear springs.
Dynamics of the chain is governed by equation
\be{EM FPU}
\begin{array}{l}
  \DS m \dot{v}_n = C(u_{n+1}-2u_n+u_{n-1}) \\
  \DS               + \alpha \Bigl((u_{n+1}-u_n)^2-(u_{n}-u_{n-1})^2\Bigr),
\end{array}
\ee
where $u_n, v_n$ are displacement and velocity of the particle~$n$, $C$ is the stiffness, and $\alpha$ is a parameter characterizing nonlinearity. Periodic boundary conditions~$u_{n} = u_{n+N}$ are used.

\edited{We separate mechanical and thermal displacements of particles as follows~\cite{Krivtsov Enciclop}. By the definition, mechanical motion is associated with time evolution of mathematical expectation of particles displacements. Macroscopic displacement field~$u(x, t)$, corresponding to mechanical motion, is such that 
\be{mech} 
   u(na, t) = \av{u_n},
\ee
where $a$ is the lattice constant, $\av{...}$ stands for mathematical expectation~(in computer simulations it is replaced by average over realizations with different initial condition). Macroscopic mechanical energy is calculated using the displacement field~$u(x,t)$~(see formula~\eq{en1}).}

\edited{The thermal motion is defined as the difference between the total displacement and the mechanical one. Then thermal displacements, $\tilde{u}_n$,  are calculated as
\be{therm}
\tilde{u}_n = u_n - \av{u_n}.
\ee 
Note that in contrast to mechanical displacements, the thermal displacements are random. 
Similar separation is carried out for particle velocities.
Then kinetic temperature, $T_n$, of particle~$n$ is defined as
\be{temp}
k_B T_n = m \av{\tilde{v}_n^2},
\ee
where $k_B$ is the Boltzmann constant.
\edited{Note that proper choice of definition for temperature in nonequilibrium systems remains an open question. We use definition~\eq{temp}, because it has clear physical meaning~(kinetic energy per particle) and it is easy to compute. A comprehensive  discussion of various definitions of temperature is presented e.g. in paper~\cite{Hoover temp}.}}

Using definitions~\eq{mech}, \eq{therm}, \eq{temp} we introduce initial conditions, corresponding to spatially sinusoidal kinetic temperature profile, zero initial fluxes,
and no macroscopic mechanical motions
\be{IC part}
\begin{array}{l}
   \DS u_{n} = 0, \quad v_n = \xi_n\sqrt{\frac{2k_B}{m}\(T_b + \Delta T \sin \frac{2\pi n}{N}\)},
   \\[4mm]
   \DS \av{\xi_n}=0, \qquad  \av{\xi_n^2}=1,
\end{array}
\ee
where  $\xi_n$ are uncorrelated random numbers with zero mean and unit variance;  $T_b$ is the average~(background) temperature; $\Delta T$ is an amplitude of the initial temperature profile. Note that in real experiments, similar initial conditions can be realized in the framework of transient thermal grating technique~\cite{Johnson Temp Grat, Rogers Therm Grating, Huberman graphite experiment}.

Heat transfer in harmonic, $\alpha\beta$-FPUT, and $\beta$-FPUT chains with initial conditions~\eq{IC part}
was investigated in papers~\cite{Sokolov 2019 PRE, Gendelman 2010 nonstat, Zhao Wang sin}.
However in these works thermoelastic effects and thermalization were not considered.

Note that initial conditions used in the original FPUT problem~\cite{FPU} significantly differ from
initial conditions~\eq{IC part}. In paper~\cite{FPU}, deterministic initial conditions,
corresponding to excitation of the first mode of mechanical vibrations at zero temperature were considered.
In contrast, initial conditions~\eq{IC part} are stochastic. Temperature and thermal energy of the
chain are finite, while initial mechanical motions are absent. Consequences of this difference in initial conditions are discussed 
in section~\ref{sect thermal}.


\section{Ballistic resonance (theory)}\label{sect theory}
We present a continuum model, describing macroscopic linear thermoelasticity of
the $\alpha$-FPUT chain~\eq{EM FPU}. Using the model, we describe a new resonance phenomenon.

\subsection{Sinusoidal initial temperature profile}

We assume that macroscopic  mechanical motions of the chain are described by the equation of linear thermoelasticity~\cite{Krivtsov Enciclop}, while  behavior of temperature~(heat transfer) is described by the ballistic heat equation~\cite{Krivtsov DAN 2015, Krivtsov 2019 ballistic}.
Conversion of mechanical energy to thermal energy is neglected. Then macroscopic behavior of the chain in continuum limit is described by equations
\be{thermoelast}
\begin{array}{l}
  \DS \ddot{u} = c_s^2 \Bigl(u'' - \beta T'\Bigr),
  \qquad 
  \end{array}
\ee
\be{ballist}
\begin{array}{l}
	\DS 
	\qquad \ddot{T} +\frac{1}{t}\dot{T} = c_s^2 T''.
\end{array}
\ee
Here $u(x,t)$ is displacement field; $T(x,t)$ is temperature field; prime stands for spatial derivative; $c_s = \sqrt{E/\rho}$ is speed of sound; $E$ is Young's modulus; $\rho$ is density; $\beta$ is thermal expansion coefficient. The relation between macroscopic and microscopic parameters of the chain are given by formulae~\eq{mech}, \eq{temp}, and \eq{cont vs discr}. 

\edited{We note that both macroscopic equations~\eq{thermoelast} and \eq{ballist} are derived from equations of motion~\eq{EM FPU}. Anharmonic effects are taken into account only in equation~\eq{thermoelast}. Equation~\eq{ballist} is derived using harmonic approximation~\cite{Krivtsov DAN 2015, Krivtsov 2019 ballistic} and therefore it corresponds to purely ballistic heat transport regime. However it is shown below that this equation describes evolution of temperature with acceptable accuracy, at least for some time, depending on nonlinearity parameter~$\alpha a/C$~(see e.g. figure~\ref{temperature FPU}).}

Periodic boundary conditions and the following initial conditions,
corresponding to microscopic conditions~\eq{IC part}, are used
\be{IC cont}
  T = T_b + \Delta T \sin(\lambda x), \quad \dot{T}=0,  \quad u = 0, \quad v = 0,
\ee
where $\lambda = 2\pi/L$, $L$ is the chain length. 

Solution of the ballistic heat equation~\eq{ballist} with initial conditions~\eq{IC cont}
 has the form~\cite{Krivtsov DAN 2015}
\be{temp harm}
   T = T_b + A(t)  \sin(\lambda x),
   \qquad
   A(t) = \Delta T J_0\!\(\omega t\),
\ee
where $\omega = \lambda c_s$, and $J_0$ is the Bessel function of the first kind. Formula~\eq{temp harm} shows that temperature oscillates in time~(see figure~\ref{temperature FPU}).

Substituting expression~\eq{temp harm} into dynamics equation~\eq{thermoelast}, we obtain:
\be{eq}
\ddot{u} = c_s^2 u''
-
\lambda c_s^2 \beta\Delta T J_0\!\(\omega t\)  \cos(\lambda x).
\ee
It is seen that the temperature acts as an ``external force'', exciting the first normal mode of mechanical vibrations. \edited{From properties of the Bessel function~$J_0$ it follows that the external force oscillates with  frequency~$\omega$ and decays as~$1/\sqrt{t}$. Note that the frequency coincides with the first eigenfrequency of mechanical vibrations.}

Solution of equation~\eq{eq} yields an exact expression for displacements
\be{u exact}
  u(x, t) = z(t) \cos(\lambda x),
  \qquad
  z(t) =
  -\beta \Delta T \,\omega t J_1(\omega t)/\lambda.
\ee
At large times~($\omega t \rightarrow \infty$), amplitude
of displacements, $z$, have the following asymptotic behavior
\be{z ass}
  z(t) \approx  - \sqrt{\frac{2}{\pi}}\frac{\beta \Delta T}{\lambda}  \sqrt{\omega t} \cos\(\omega t - 3\pi/4\).
\ee
Formula~\eq{z ass} shows that the amplitude growth in time as~$\sqrt{t}$.
Corresponding mechanical energy, $\mathcal{E}$, is calculated via
\be{en1}
  \mathcal{E} = \frac{1}{2L}\int_0^L\!\!\(\rho v^2 + E u'^2 \){\rm d}x  =
  \mathcal{E}_* \omega^2 t^2 \Bigl[J_0\!^2(\omega t) + J_1^2(\omega t)\Bigr],
\ee
%
%
%
\edited{where~$\mathcal{E}_* = E \beta^2 \Delta T^2/4$ is proportional to the potential energy of the system due to thermal expansion in the case of uniform temperature profile~$\Delta T$.}  At large times, the energy grows linearly, i.e.~$\mathcal{E} \approx 2\mathcal{E}_* \omega t/\pi$.

Thus the coincidence of a frequency of temperature oscillations with the first eigenfrequency of the chain leads to excitation of mechanical vibrations with growing amplitude.
This new phenomenon is referred to as ballistic resonance.

\subsection{Periodic initial temperature profile}
\edited{We generalize presented results for the case of an arbitrary periodic initial temperature profile~$T_0(x)=T_0(x+L)$.
Fourier series expansion of the profile yields:
\be{Fourier}
\begin{array}{l}
	  \DS T_0 = \frac{a_0}{2} + \sum_{k=1}^{\infty} \Bigl[a_k \cos\(\lambda_k x\) + b_k \sin\(\lambda_k x\)\Bigr], \quad 
	\lambda_k= k \lambda, \\[4mm]
	\DS a_k = \frac{2}{L}\int_{0}^{L} T_0(x)\cos(\lambda_k x) {\rm d}x,
	\\[4mm] 
	\DS b_k = \frac{2}{L}\int_{0}^{L} T_0(x)\sin(\lambda_k x) {\rm d}x.
\end{array}
\ee	
Solution of equations~\eq{thermoelast}, \eq{ballist}, corresponding to temperature  profile~\eq{Fourier}, is derived using formula~\eq{u exact} and the superposition principle:
\be{u mult}
  u = \beta c_s t \sum_{k=1}^{\infty} J_1(\omega_k t) \Bigl[a_k \sin\(\lambda_k x\) - b_k \cos\(\lambda_k x\)\Bigr],
\ee
where $\omega_k = k \omega = k \lambda c_s$. Formula~\eq{u mult} shows that all eigenmodes, included in the expansion of function~$T_0$, resonate. 
Amplitudes of these modes grow in time as~$\sqrt{t}$. However since Fourier coefficients $a_k$, $b_k$ decay with increasing~$k$~\footnote{This statement holds at least for all differentiable functions.} 
then the main contribution to growth of displacements is given by long wavelength harmonics~(small $k$).}

\edited{Thus the ballistic resonance occurs for any periodic distribution of the initial temperature. }

\section{Ballistic resonance (numerical results)}
\edited{We compare predictions of the presented continuum theory with results of numerical solution of discrete equations of motion~\eq{EM FPU}.}

\edited{Macroscopic length, density, Young' modulus, speed of sound, and the thermal expansion coefficient
 are related to the micro parameters of the chain as~\cite{Krivtsov Enciclop}
\be{cont vs discr}
\begin{array}{l}
	\DS L = N a,
	\qquad
	\rho = \frac{m}{a},
	\qquad
	E = C a,
	\qquad
	c_s = a\sqrt{\frac{C}{m}},\\[4mm]
	\DS
	\beta = -\frac{\alpha k_B}{a C^2}.
\end{array}
\ee
Here thermal expansion coefficient is calculated as~$\beta = \Gamma k_B /(E a)$, where~$\Gamma = -\alpha a/C$
is the Gruneisen parameter~(see e.g.~\cite{Krivtsov Enciclop, Krivtsov EOS 1D}).}

\edited{To compute macroscopic mechanical characteristics~(e.g. $z$ and $\mathcal{E}$),
we consider~$N_r$ realizations of the chain~\eq{EM FPU} with random initial conditions~\eq{IC part}. For each realization, equations of motion are solved numerically using the 4th order symplectic integrator~\cite{Candy Rozmus integrator}
with optimized parameters~\cite{McLachlan Atela integrator}.
In our simulations the total energy is conserved with an accuracy of order of~$0.001\%$.}

The amplitude of mechanical vibrations, $z$,
and mechanical energy, $\mathcal{E}$, were computed via
\be{comp quantities}
\begin{array}{l}
  \DS z \approx  -\frac{1}{\pi}\sum_{n=0}^{N-2}  \av{u_{n+1}-u_n}_r \sin\frac{2\pi n}{N},
  \\[4mm]
   \DS \dot{z} \approx \frac{2}{N-1}\sum_{n=0}^{N-1} \av{v_n}_r \cos\frac{2\pi n}{N}, \quad
  \mathcal{E} = \frac{m}{4 a} \(\dot{z}^2 + \omega^2 z^2\).
\end{array}
\ee
Here $\av{...}_r$ represent averaging over the realizations.

To investigate the influence of anharmonic effects, we fix the background temperature~$T_b$
and change parameter of nonlinearity~$\alpha a/C$  in the interval~$[-1; 0]$. For the remaining parameters, the following values were used:
\be{}
\begin{array}{l}
	\DS \Delta t = 0.05 \tau_*, \quad t_{max} = 1.4 \cdot 10^4 \tau_*, \quad \tau_* = 2\pi\sqrt{m/C}, 
  \\[4mm]
  \DS  \Delta T = 0.5 T_b,\quad  
  \DS  v_0 = 0.1 c_s, \quad N = 10^3, \quad  N_{r} = 10^4.
\end{array}
\ee
Here~$v_0$ is the amplitude of random initial velocities, corresponding to background temperature~$T_b$.

\begin{figure}
\begin{center}
\includegraphics[width=.5\textwidth]{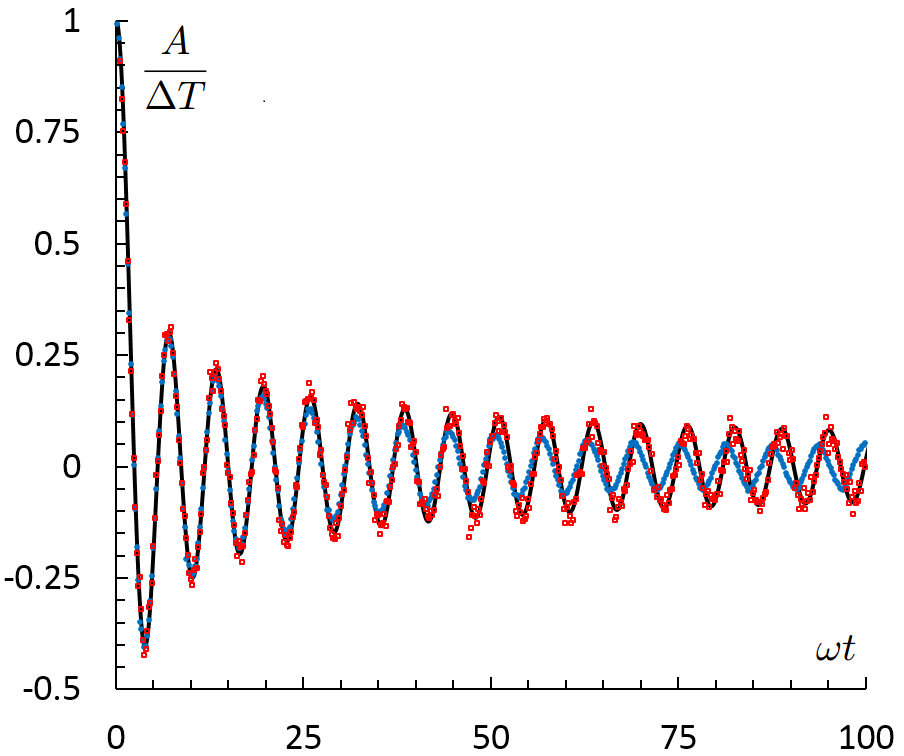}
\caption{Oscillations of temperature caused by quasiballistic heat transport.
Analytical solution~\eq{temp harm}~(line), and results of numerical integration of
equations of motion for~$\alpha a/C =-0.25$~(red squares) and $-1$~(blue circles).}
\label{temperature FPU}
\end{center}
\end{figure}

We consider behavior of kinetic temperature of the chain. Analytical solution~\eq{temp harm} suggests
that the temperature
profile remains sinusoidal. Using this fact, we compute the amplitude,~$A(t)$,
of the temperature profile~(see formula~\eq{temp harm}). In numerical simulations,
the temperature is calculated
by definition~\eq{temp}. It can be shown that the contribution of~$\av{v_n}$
to the amplitude~$A$ is negligible.
Therefore in formula~\eq{temp} the total particle velocity~$v_n$ is used instead of~$\tilde{v}_n$.
The amplitude for $\alpha a/C = -0.25$ and $\alpha a/C = -1$ is shown in figure~\ref{temperature FPU}. For both values of $\alpha$ the temperature oscillates in time. These oscillations are responsible for the ballistic resonance. For $\alpha a/C=-0.25$ the oscillations are described by the analytical solution~\eq{temp harm} with high accuracy. Deviations from the analytical solution for $\alpha a/C= -1$ are caused by anharmonic effects,
neglected in derivation of the ballistic heat equation~\eq{ballist}.


%
\begin{figure}
	\begin{center}
	\includegraphics[width=.5\textwidth]{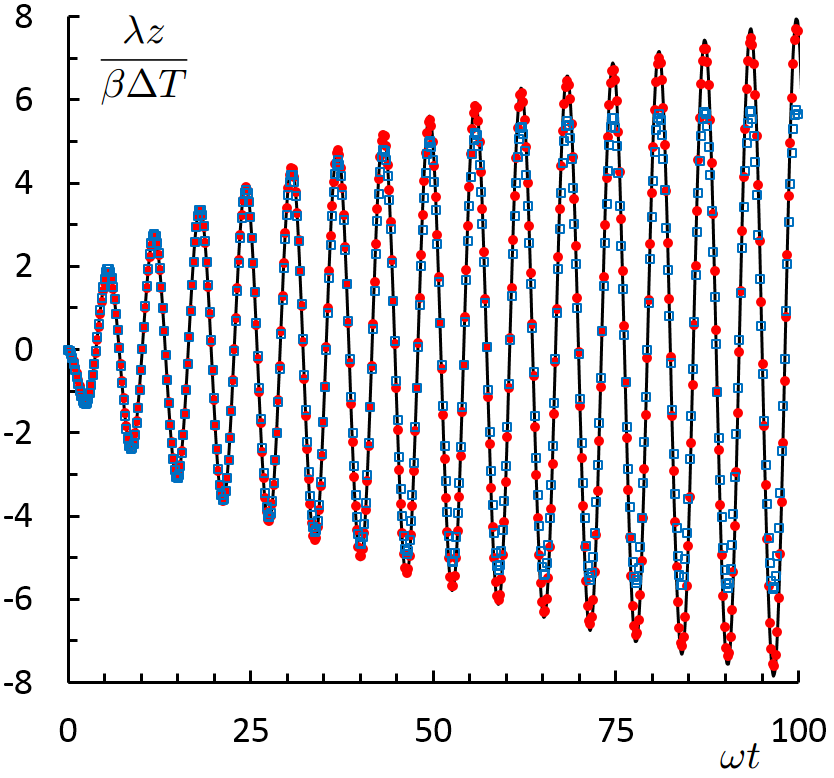}
	\caption{Growth of amplitude of mechanical vibrations due to ballistic resonance.
		Analytical solution~\eq{u exact}~(solid line) and numerical results for~$\alpha a/C=-0.25$~(circles) and $-1$~(squares).}
	\label{amplitude short time}
	\end{center}
\end{figure}
The time dependence of amplitude of mechanical vibrations,~$z(t)$,
is presented in figure~\ref{amplitude short time}. It is seen that the amplitude grows in time.
 Initial growth is accurately described by the analytical solution~\eq{u exact}.
 Over time, an analytical solution deviates from the numerical solution.
The rate of deviation increases with increasing absolute value of the nonlinearity coefficient~$\alpha a/C$.

\begin{figure}
	\begin{center}
	\includegraphics[width=.5\textwidth]{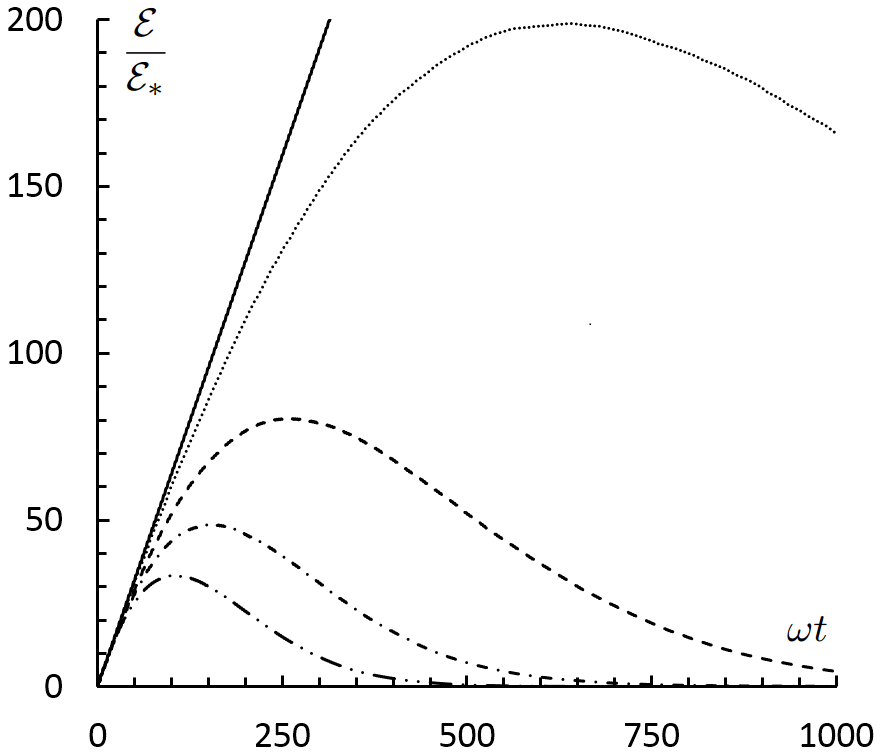}
	\caption{Time dependence of mechanical energy~$\mathcal{E}$. Analytical solution~\eq{en1}~(solid line) and numerical results for~$\alpha a/C=-0.25$ (dots), $-0.5$ (dashed line), $-0.75$ (dash-dotted line), and $-1$~(dashed double dotted line).}
	\label{energy large times}
	\end{center}
\end{figure}

The growth of amplitude of mechanical vibrations is due to the partial
conversion of thermal energy to mechanical energy.
The conversion is clearly seen in figure~\ref{energy large times}.
The figure shows that initially the mechanical energy grows with time
as predicted by analytical solution~\eq{en1}.
Since the total energy of the system is conserved, the growth of mechanical energy
is associated with a decrease of thermal energy.

\edited{
Thus the phenomenon of ballistic resonance, predicted by continuum theory~(equations~\eq{thermoelast}, \eq{ballist}), is observed in direct numerical simulations.}

\section{Thermalization (numerical results)}\label{sect thermal}
\begin{figure}
	\begin{center}
		\includegraphics[width=.5\textwidth]{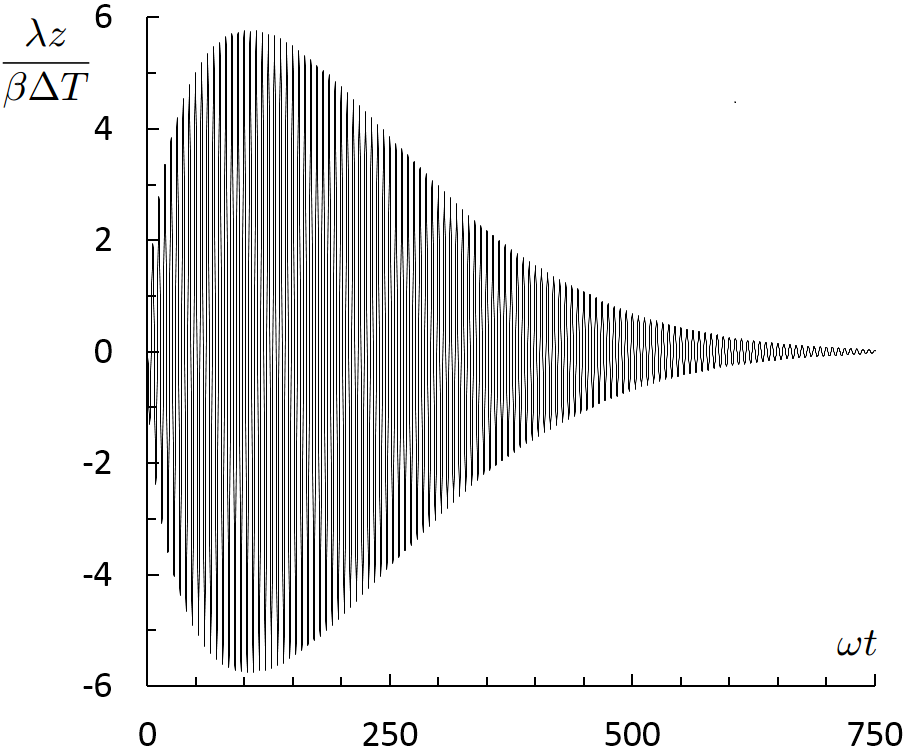}
		\caption{Decay of amplitude of mechanical vibrations at large times~(numerical results for~$\alpha a/C = -1$).}
		\label{amplitude large times}
	\end{center}
\end{figure}

 Numerical simulations show that mechanical vibrations, excited by the ballistic resonance,
decay in time~(see figure~\ref{amplitude large times}).
The decay is caused by thermalization, i.e. conversion of mechanical energy to thermal energy. This process is not covered by our continuum model~\eq{thermoelast}, \eq{ballist}.
Decay of mechanical  energy
is clearly seen in figure~\ref{energy large times}.
The energy reaches the maximum value,
depending on nonlinearity coefficient~$\alpha a/C$, and then monotonically tends to zero.

\begin{figure}
\begin{center}
\includegraphics[width=.5\textwidth]{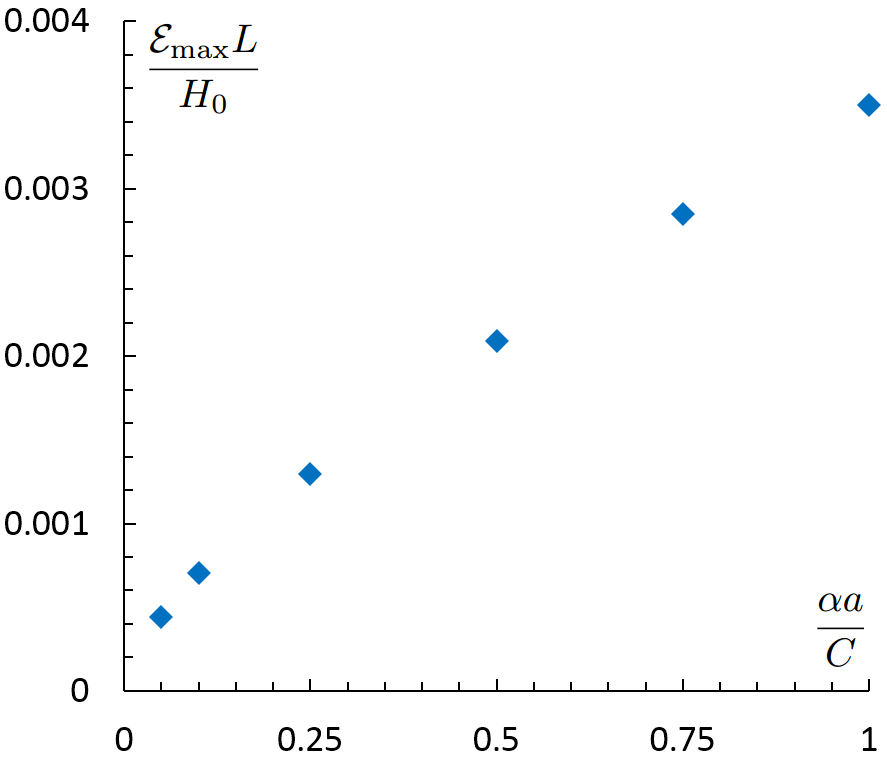}
\caption{Maximum mechanical energy, $\mathcal{E}_{max}$, excited by the ballistic resonance for different values of nonlinearity coefficient~$\alpha a/C$. Here $H_0$ is the total energy of the chain.}
\label{max mech}
\end{center}
\end{figure}

The maximum mechanical energy, $\mathcal{E}_{max}$,
excited by the ballistic resonance, normalized by the total energy of the chain, $H_0$,
is shown in figure~\ref{max mech}. $\mathcal{E}_{max}$ increases with
increasing absolute value of the nonlinearity coefficient~$\alpha a/C$.
For all considered values of~$\alpha a/C$,
the maximum mechanical energy is several orders of magnitude
less than the thermal energy. We note that in the original FPUT problem~\cite{FPU}
the situation is opposite, because thermal energy is equal to zero,
while the mechanical energy is finite.

Thus under considered initial conditions, the mechanical energy of the chain is
monotonically converted into thermal energy.
This fact is in agreement with results of numerical simulations,
carried out in paper~\cite{Tsvetkov Krivtsov} with different initial conditions~\footnote{In paper~\cite{Tsvetkov Krivtsov}, decay of the first normal modes of mechanical vibrations at finite background temperature was considered.}. 
Therefore the FPUT recurrence paradox is eliminated by finite thermal motion.

\section{Conclusions}
\edited{We have shown that excitation of a periodic~(e.g. sinusoidal) initial temperature profile in $\alpha$-FPUT chain
leads to mechanical
resonance. Temperature of the chain oscillates in time due to quasiballistic~(wave) nature
of heat transport. Nonuniform distribution of temperature causes thermal expansion,
which plays the role of a periodic force, exciting macroscopic mechanical vibration.
Frequency of this ``force'' coincide with eigenfrequency of mechanical vibrations leading
to the resonance. A continuum model of the ballistic resonance is developed.
It is shown that predictions of the model are in good agreement with results of numerical simulations.
Note that in contrast to conventional
mechanical resonance, the ballistic resonance occurs in the closed system without any external excitation.}

We expect that ballistic resonance may be observed in two- and three-dimensional crystals. However since in $d$-dimensional case sinusoidal temperature profile decays as~$t^{d/2}$~\cite{Kuzkin 2019 CMAT ballistic}
then the effect of ballistic resonance will be weaker.

It was shown that mechanical vibrations, caused by the ballistic resonance,
decay monotonically. Therefore the well-known FPUT recurrence paradox is eliminated by
adding a finite thermal motion. In our simulations the energy of mechanical vibrations is much smaller than
the thermal energy. This fact appears to be a necessary condition for monotonic decay
of mechanical vibrations. However further work is needed to prove this statement rigorously.

We are deeply grateful to H.E. Huppert, W.G. Hoover, E.A. Ivanova, S.N. Gavrilov, and
M.L. Kachanov for useful discussions. Valuable comments of the reviewers are highly appreciated. V.A. Kuzkin acknowledges support of the Russian Foundation for Basic Research~(grant No.~20-37-70058).

\end{document}